\documentclass[magazine]{IEEEtran}
\usepackage{amsmath,amssymb,amsfonts}
\usepackage{algorithmic}
\usepackage{array}
\usepackage[caption=false,font=normalsize,labelfont=sf,textfont=sf]{subfig}
\usepackage{textcomp}
\usepackage{stfloats}
\usepackage{url}
\usepackage{verbatim}
\usepackage{graphicx}
\def\BibTeX{{\rm B\kern-.05em{\sc i\kern-.025em b}\kern-.08em
		T\kern-.1667em\lower.7ex\hbox{E}\kern-.125emX}}
\usepackage{balance}

\usepackage{cite}
\usepackage{xcolor}

\usepackage{subfloat}
\usepackage{epstopdf}

\begin{document}
    \include{header}
    \title{Integrated Sensing, Communication, and Powering (ISCAP): Towards Multi-functional 6G Wireless Networks}

    \author{Yilong Chen, Zixiang Ren, Jie Xu, Yong Zeng, Derrick Wing Kwan Ng, and Shuguang Cui \\ 
        \thanks{Y. Chen is with the Future Network of Intelligence Institute (FNii) and the School of Science and Engineering (SSE), The Chinese University of Hong Kong (Shenzhen), Shenzhen, China (e-mail: yilongchen@link.cuhk.edu.cn).}
        \thanks{J. Xu and S. Cui are with the SSE and the FNii, The Chinese University of Hong Kong (Shenzhen), Shenzhen, China (e-mail: xujie@cuhk.edu.cn; shuguangcui@cuhk.edu.cn).}
        \thanks{Z. Ren is with the Key Laboratory of Wireless-Optical Communications, Chinese Academy of Sciences, School of Information Science and Technology, University of Science and Technology of China, Hefei 230027, China, and also with the FNii, The Chinese University of Hong Kong (Shenzhen), Shenzhen 518172, China (e-mail: rzx66@mail.ustc.edu.cn).}
        \thanks{Y. Zeng is with the National Mobile Communications Research Laboratory, Southeast University, Nanjing, China. He is also with the Purple Mountain Laboratories, Nanjing, China (e-mail: yong\_zeng@seu.edu.cn).}
        \thanks{D. W. K. Ng is with School of Electrical Engineering and Telecommunications, the University of New South Wales, Sydney, Australia (e-mail: w.k.ng@unsw.edu.au).}
	\thanks{J. Xu is the corresponding author.}
        \thanks{Y. Chen and Z. Ren are the co-first authors.}
    }
    
    \maketitle
    
\begin{abstract}
This article presents a novel multi-functional system for sixth-generation (6G) wireless network with integrated sensing, communication, and powering (ISCAP), which unifies integrated sensing and communication (ISAC) and wireless information and power transfer (WIPT) techniques. The multi-functional ISCAP network promises to enhance resource utilization efficiency, reduce network costs, and improve overall performance through versatile operational modes. Specifically, a multi-functional base station (BS) can enable multi-functional transmissions, by exploiting the same radio signals to perform target/environment sensing, wireless communication, and wireless power transfer (WPT), simultaneously. Besides, the three functions can be intelligently coordinated to pursue mutual benefits, i.e., wireless sensing can be leveraged to enable light-training or even training-free WIPT by providing side-channel information, and the BS can utilize WPT to wirelessly charge low-power devices for ensuring sustainable ISAC. Furthermore, multiple multi-functional BSs can cooperate in both transmission and reception phases for efficient interference management, multi-static sensing, and distributed energy beamforming. For these operational modes, we discuss the technical challenges and potential solutions, particularly focusing on the fundamental performance tradeoff limits, transmission protocol design, as well as waveform and beamforming optimization. Finally, interesting research directions are identified.
\end{abstract}

\section{Introduction}


Sixth-generation (6G) mobile networks are envisioned to support various intelligent applications such as smart cities, smart homes, and smart manufacturing, by providing a peak data rate of 1 Tbps, a user-experienced data rate of 10-100 Gbps, a localization accuracy of 1 cm indoors and 50 cm outdoors, and a battery lifetime of up to 20 years for low-power Internet-of-Things (IoT) devices \cite{tong20226g}.
Towards this end, 6G networks need to incorporate a massive number of wireless devices, such as smartphones, sensors, IoT devices, and network-connected robots for achieving integrated sensing and communications (ISAC), which has been identified by IMT-2030\footnote{For more details, please refer to https://www.itu.int/en/ITU-R/study-groups/rsg5/rwp5d/imt-2030.} as one of the core usage scenarios for 6G.
On the other hand, to support the sustainable operation of massive devices (especially low-power IoT devices), wireless power transfer (WPT) can be integrated into 6G networks to enable wireless information and power transfer (WIPT) \cite{clerckx2018fundamentals}. By unifying ISAC and WIPT, 6G is envisioned to become a multi-functional wireless network with integrated sensing, communication, and powering (ISCAP) at the air interface. 

The multi-functional ISCAP network is expected to offer a plethora of applications. In such a network, inter-connected base stations (BSs) and wireless devices may act as giant networked sensors to monitor the environment. 
The distributed sensing data are then wirelessly transmitted among separate edge servers at BSs, which can process such data via advanced signal processing or artificial intelligence (AI) algorithms and take inference or control actions accordingly. 
BSs in this network can also reuse wireless sensing and communication signals for a triple role of WPT, which wirelessly charges low-power devices to support their ISAC functionalities sustainably. 
For ISCAP operations, recent advancements in millimeter wave (mmWave)/terahertz (THz) and extremely large-scale antenna arrays \cite{lu2023tutorial} are particularly beneficial.
By exploiting the wide bandwidth at mmWave/THz bands and high beam directivity of large antenna arrays, ISCAP systems can significantly enhance the sensing resolution and accuracy, substantially increase the communication data rate, lower transmission latency, and improve the energy transfer efficiency by combating the severe signal path loss. 

Different from conventional sole-functional wireless networks, multi-functional ISCAP networks offer several advantages. 
First, the triple use of spectrum/energy resources and hardware platforms significantly enhances resource utilization efficiency and concurrently reduces both capital and operational costs. Next, the joint sensing-communication-powering optimization in ISCAP systems can significantly improve the system performance by effectively harnessing interference. Furthermore, the sensing-communication-powering integration allows mutual benefits between different functions. For instance, wireless sensing can provide real-time environment information and channel state information (CSI) to facilitate WIPT, wireless powering can provide a convenient energy supply for low-power IoT devices to support their sustainable sensing and communications, and communications among distributed nodes can enable their cooperative transmission to enhance both wireless sensing and powering capabilities. 
In general, the implementation of such multi-functional ISCAP networks can be categorized into the following four distinct modes.

\begin{itemize}
\item \textbf{Simultaneous multi-functional transmission:} Multi-functional BS utilizes the same radio signals for sensing targets, sending messages to information receivers (IRs), and charging energy receivers (ERs) concurrently. 
This mode significantly enhances the spectrum utilization efficiency by realizing the three functions at the same time over the same frequency bands.

\item \textbf{Sensing-assisted WIPT:} Multi-functional BS adopts wireless sensing to acquire environmental information (e.g., locations of scatterers), which is then leveraged to facilitate the CSI acquisition for efficient WIPT.

\item \textbf{Wireless powered ISAC:} Multi-functional BS employs WPT in the downlink to charge low-power IoT devices, which can then utilize the harvested energy to perform ISAC sustainably. 

\item \textbf{Multi-BS cooperation for multi-functional operation:} Distributed multi-functional BSs are coordinated in signal transmission and reception for fulfilling the multiple functions. With proper data/CSI sharing and time-frequency synchronization, these BSs are capable of performing multi-static or distributed multiple-input multiple-output (MIMO) sensing \cite{liu2023seventy}, coordinated beamforming or cell-free MIMO transmission \cite{liu2022integrated},
and distributed energy beamforming \cite{zhang2022near}, thus enhancing the performance of large-scale networks.
\end{itemize}

This article presents an overview of multi-functional ISCAP networks, focusing on the above four operational modes. For each mode, we present the representative application scenarios and discuss the technical challenges and potential solutions from the perspectives of fundamental performance limits, waveform and beamforming design, and interference management. We also highlight some interesting research directions to motivate future research.

It is worth noting that there is a prior work \cite{li2023integrating} investigating multi-functional networks by considering the simultaneous multi-functional transmission and wireless-powered sensing and communications. By contrast, this article presents a more comprehensive overview of the four operational modes, offering deeper technical insights.

\section{Simultaneous Multi-functional Transmission}

\begin{figure}[tb]
    \centering {\includegraphics[width=0.45\textwidth]{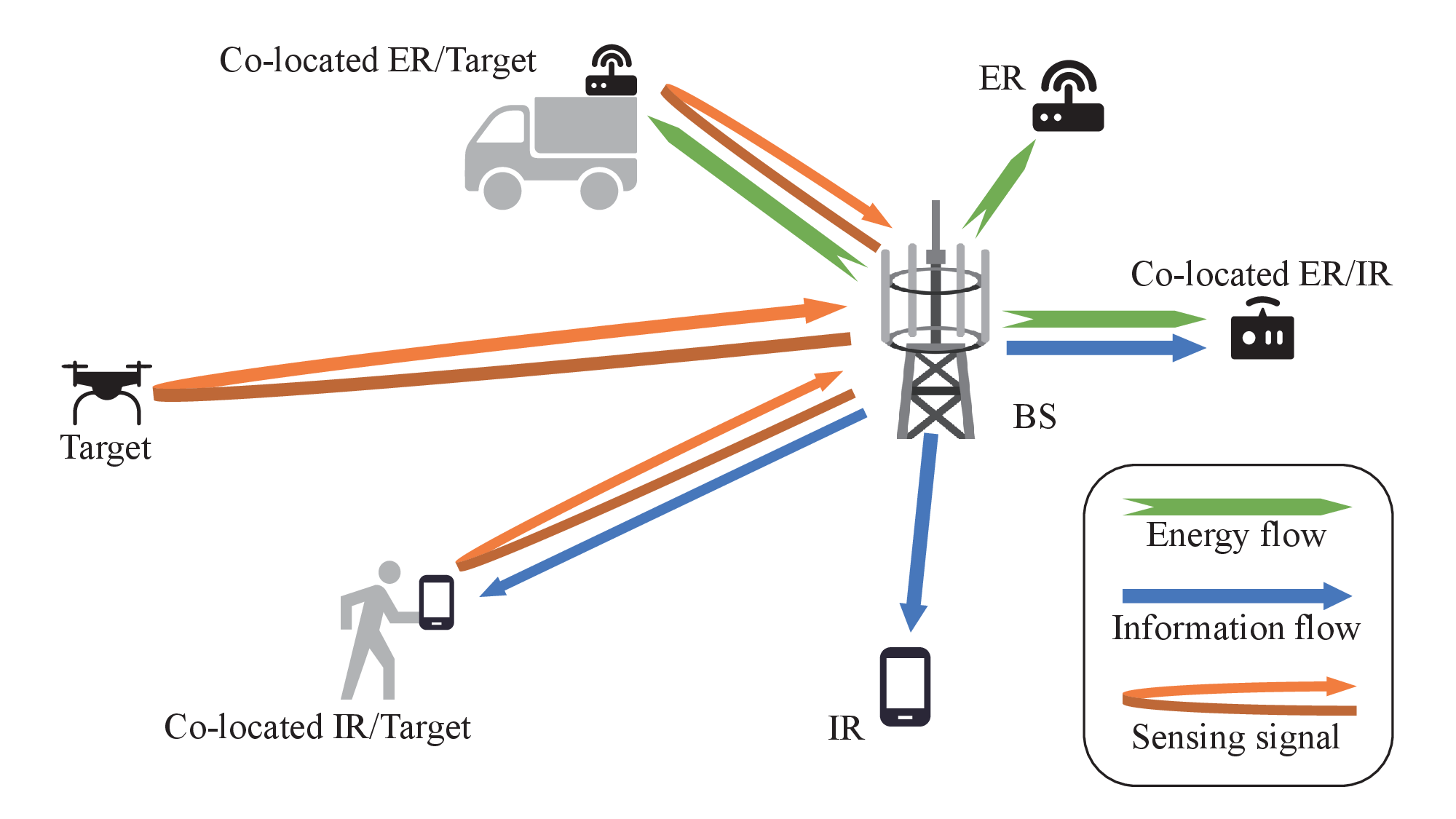}} 
    \caption{Simultaneous multi-functional transmission with one BS concurrently communicating with IRs, charging ERs, and sensing targets.}
\end{figure}
This section considers simultaneous multi-functional transmission, as illustrated in Fig. 1, in which a multi-functional BS transmits unified signals for concurrently delivering information to IRs, providing energy supply to ERs, and performing target sensing based on echo signals. In this system, the IRs, ERs, and targets can be separated at different locations, or co-located at the same location. 
This design capitalizes on the simultaneous triple functionality of radio signals, which is anticipated to significantly enhance the utilization efficiency of the scarce spectrum and hardware resources.

The practical implementation of simultaneous multi-functional transmission presents new technical challenges. First, the three aforementioned functions operate based on different performance indicators and adhere to different design principles. For instance, the estimation mean square error (MSE) and Cram{\'e}r-Rao bound (CRB) (corresponding to the MSE lower bound of any unbiased estimators) have been widely adopted as sensing performance measures, while channel capacity and communication data rate are employed for communication and harvested direct-current (DC) power is commonly considered for WPT. Consequently, wireless sensing, communication, and WPT follow different design principles, particularly in waveform and beamforming optimization. Consider a MIMO system as an example,
isotropic transmission is usually needed for minimizing the CRB/MSE in a MIMO wireless sensing system for extended target estimation. In contrast, eigenmode transmission (EMT) with water-filling (WF) power allocation is optimal for achieving capacity in a point-to-point MIMO communication system, while the strongest EMT is generally preferred for maximizing the received radio frequency (RF) power in a MIMO WPT system \cite{chen2022isac}. Therefore, by jointly considering these distinct design principles, how to find proper transmit strategies to optimally balance the performance tradeoff among sensing, communication, and powering poses a new challenge. 
Next, the three systems handle interference differently. 
For communication, interference is detrimental and must be mitigated through advanced signal processing techniques. Conversely, for WPT, interference can be a beneficial factor as it carries energy that can be harvested by ERs. For sensing, interference inherently contains environmental information, which, however, is only extractable provided that the signal-embedded information is known. As a result, properly managing interference across different functionalities remains another crucial technical challenge.


To explore the fundamental performance limits of simultaneous multi-functional transmission, the authors in \cite{chen2022isac} studied a basic MIMO system with a multi-antenna BS serving a multi-antenna IR, a multi-antenna ER, and a target to be sensed. 
They introduced the CRB-rate-energy (C-R-E) region to characterize the performance tradeoff among sensing, communication, and powering, which is defined as the set encompassing all simultaneously achievable C-R-E triplets. In this context, the objective is to identify efficient transmission strategies that can achieve the Pareto boundary of the C-R-E region. The boundary represents the set of points at which further improving one performance metric will inevitably lead to a compromise in another.
To achieve this, the authors proposed to optimize the transmit covariance for maximizing the communication rate while ensuring the estimation CRB requirement for target sensing and the harvested energy requirement at the ER, subject to the maximum transmit power constraint at the BS.
In particular, the obtained optimal transmit covariance in \cite{chen2022isac} follows a new EMT structure based on a composite channel matrix of communication, powering, and sensing channels, combined with a water-filling-like power allocation. Specifically, the solution exhibits a block structure, partitioned into two sections: one serving the triple roles of wireless communication, sensing, and powering, and the other exclusively for sensing and powering. 
This solution effectively unifies the optimal transmit strategies for MIMO communication, sensing, and WPT, as well as those for MIMO SWIPT and ISAC \cite{chen2022isac}. 

Fig. 2 illustrates the Pareto boundaries of the C-R-E regions for a multiple-input single-output (MISO) ISCAP system consisting of a BS equipped with a uniform linear array (ULA), an IR, and an ER each with a single antenna, as well as a point target for sensing. Suppose that there are \(6\) transmit antennas and \(16\) receive antennas at the BS with half-wavelength spacing between adjacent antennas, and assume line-of-sight (LoS) channels for sensing, communication, and WPT. 
We have the following interesting observations from Fig. 2. When the three channels become identical, the Pareto boundary of C-R-E region collapses to a point, indicating that the optimal sensing, communication, and WPT strategies are identical to optimize the three performance metrics simultaneously. Conversely, when the associated channels are orthogonal to each other, optimizing one metric (e.g., harvested energy) inevitably results in poor performances in the other two metrics (e.g., a close-to-zero rate and a high CRB). This scenario highlights the inherent conflict among the three objectives. Moreover, for correlated channels, the Pareto boundary of the C-R-E region is observed to lie between the two extreme cases with identical and orthogonal channels, showcasing a more balanced interplay among the three objectives.



While \cite{chen2022isac} provides essential insights by studying a basic setup with an IR, an ER, and a target, practical wireless networks consisting of multiple IRs, ERs, and targets need to be studied. Extending the design in \cite{chen2022isac} to such practical setups introduces further challenges. Specifically, the three-dimensional (3D) C-R-E region in \cite{chen2022isac} evolves into a multi-dimensional region, thus making the characterization of performance tradeoffs intractable. Furthermore, inter-user interference may become more complicated and more advanced multiple access designs such as capacity-achieving dirty paper coding (DPC), non-orthogonal multiple access (NOMA), and rate-splitting multiple access (RSMA) may be applied for performance enhancement, for which more sophisticated beamforming designs together with successive interference cancellation are necessary. 

The extension of simultaneous multi-functional transmission from narrowband systems to their wideband counterparts is another important topic, for which signal waveform optimization becomes crucial. It is well-established that zero-mean Gaussian distributed random sequences are generally optimal in achieving channel capacity. However, for wireless sensing, especially when the sensing duration or sensing dwell time is limited \cite{liu2023seventy},  deterministic sequences with favorable auto-correlation properties are usually preferred. Furthermore, WPT tends to favor sequences with non-zero or high absolute means \cite{zhang2023multi}. These heterogeneous waveform requirements make the optimization of simultaneous multi-functional transmission important yet difficult. The authors in \cite{zhang2023multi} investigated this problem over orthogonal frequency division multiplexing (OFDM) systems, in which the non-zero mean asymmetric Gaussian distributed signal input is considered. The input distributions or the mean and variance over different subcarriers were optimized in \cite{zhang2023multi} to maximize the harvested power at the ER, subject to constraints on the achievable rate for the IR and the average side-to-peak-lobe difference for effective target sensing. 
It is observed that the optimized input distributions exhibit distinct features under different performance requirements of the three functions, which inspires the waveform design of the multi-functional system.

\begin{figure}[tb]
    \centering {\includegraphics[width=0.45\textwidth]{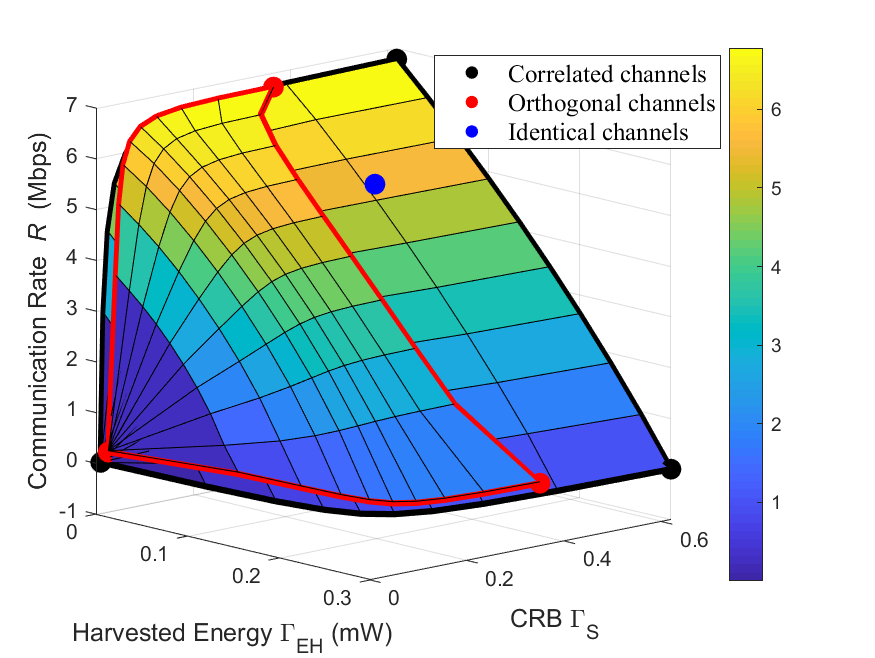}}
    \caption{The Pareto boundary of a C-R-E region example \cite{chen2022isac}.}
\end{figure}



\section{Sensing-Assisted WIPT}

The multi-functional network provides novel opportunities to exploit wireless sensing to efficiently acquire CSI for facilitating WIPT. Conventionally, CSI acquisition has been a critical yet challenging task for WIPT, especially when there are massive antennas at the BS operating in frequency-division duplex (FDD) mode. In this case, as the uplink-downlink channel reciprocity is absent, the BS sends pilots in the forward link, based on which each IR/ER estimates its associated CSI and then feeds it back to the BS after proper quantization and compression.
The process of training and feedback introduces significant overheads that can adversely burden and compromise WIPT performance. 

Leveraging wireless sensing information to assist CSI acquisition has emerged as a viable solution to enhance the performance of both information and energy transmission. In this approach, the BS explores its sensing capabilities to estimate relevant information about ERs/IRs and environmental scatterers, which is then exploited to facilitate the acquisition of CSI with significantly reduced training and feedback overheads. The rationale is that downlink communication and mono-static sensing often interact with the same environmental scatterers. A general scenario for sensing-assisted WIPT scenario is depicted in Fig. 3, where multiple detectable scatterers in the environment contribute to the channel paths. The BS may utilize its sensing capabilities to sense the environment and reconstruct the CSI leveraging the feedback path coefficients.

\begin{figure}[tb]
    \centering {\includegraphics[width=0.45\textwidth]{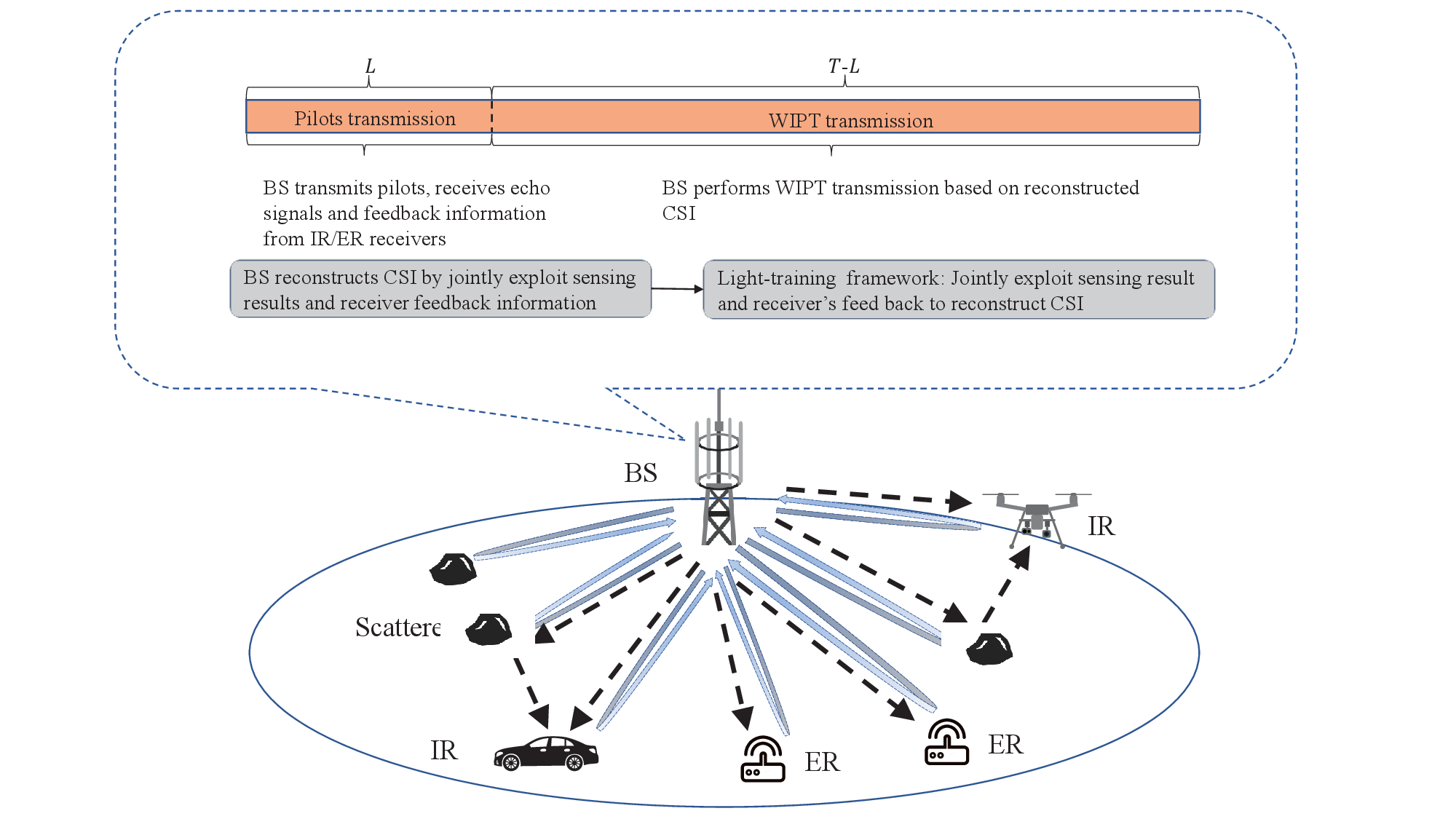}} 
    \caption{Illustration of sensing-assisted WIPT.}
\end{figure}

Generally speaking, sensing-assisted WIPT can be implemented through a two-stage downlink transmission protocol, as depicted in Fig. 3. The transmission interval $T$ is divided into a pilot transmission interval $L$ and a data/energy transmission interval $T-L$. During the pilot transmission interval, the BS sends downlink pilots to ERs/IRs while concurrently receiving echo pilot signals to sense the presence and characteristics of surrounding scatterers. Subsequently, the IR/ER receiver sends feedback information concerning its received pilot signals back to the BS transmitter. Based on the received pilot signal echoes, the BS can apply various signal processing methods to extract environmental parameters such as the angles and distances of scatterers. By combining the extracted parameters together with receiver feedback information from ERs/IRs, the BS can successfully reconstruct the CSI for facilitating the WIPT in the second interval. For example, the wireless channel in this context can be efficiently conceptualized as a multi-path channel influenced by detectable scatterers. In such environments, the BS can exploit the pilot signal echoes via wireless sensing to extract the scatterer information for each path. Moreover, the BS can also leverage the receiver feedback information to estimate the coefficient information for each path, which thus leads to effectively reconstructed CSI. Compared to the traditional downlink channel training methods, this operation can substantially reduce the required training overhead thanks to the assistance of wireless sensing,  which is thus termed a light-training framework \cite{Ren2023SensingAssistedSC}.  Furthermore, in special LoS scenarios, the light-training approach could further evolve into a training-free framework. In this case, the BS can precisely shape transmit beams towards directions identified through wireless sensing without requiring feedback information from ERs/IRs.

In particular, \cite{Ren2023SensingAssistedSC} investigated the sensing-assisted CSI acquisition problem in massive antenna systems under this light-training framework. It operates on the assumption that the number of scatterers is significantly fewer than the number of antennas at the BS side, resulting in rich channel sparsity. The wireless channel, in this context, comprises multiple paths associated with scatterers detectable through wireless sensing. Leveraging this information, the BS identifies the sparse basis from the sensed scatterers and proceeds to reconstruct the wireless channel by exploiting receiver feedback information based on advanced compressive sensing (CS) algorithms.
\begin{figure}[tb]
    \centering {\includegraphics[width=0.45\textwidth]{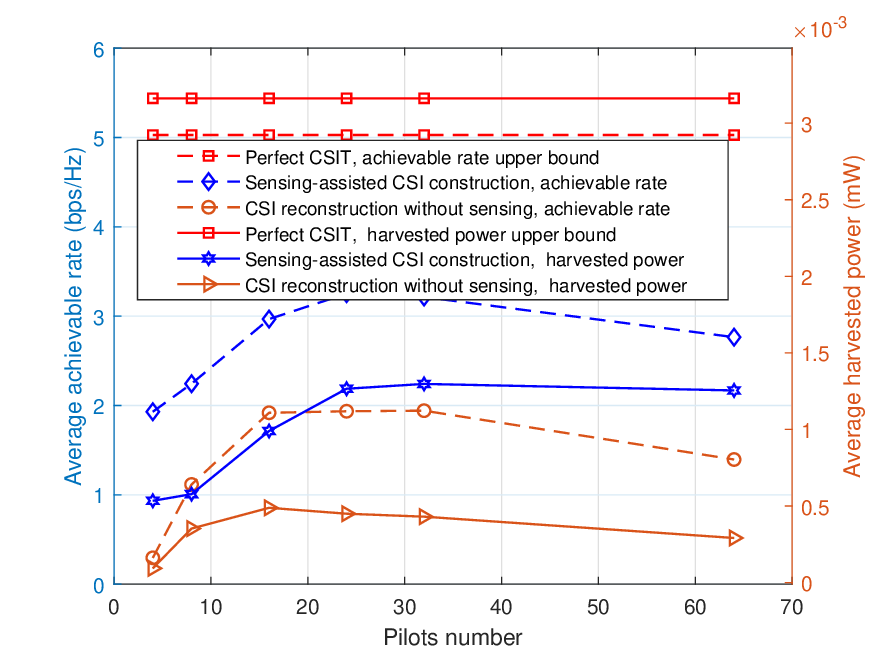}} \caption{Average achievable rate and harvested power for WIPT versus pilots number.}
\end{figure} Fig. 4 illustrates numerical results showcasing the average achievable rate and the average harvested power versus the number of pilots, based on the setup in \cite{Ren2023SensingAssistedSC}. With a setup of \(64\) antennas, an IR, an ER, and six environmental scatterers (with four contributing to the channels), the results show a significant improvement in CSI reconstruction with sensing assistance compared with traditional counterparts without sensing. This underscores the effectiveness of the proposed sensing-assisted approach, yielding tangible benefits in terms of enhanced average achievable rate and harvested power. The observed trend reveals that the achievable rate/harvested power experiences an initial increase followed by a subsequent decline as the pilot length increases. This phenomenon is attributed to the fact that more pilots can contribute to more accurate channel estimation, but also lead to a reduction in the block length available for information/energy transmission. Consequently, it is important to optimally design the pilot duration $L$ to balance the allocated resources between the two stages for optimizing the overall transmission performance.






\section{Wireless Powered ISAC}



\begin{figure}[tb]
    \centering {\includegraphics[width=0.45\textwidth]{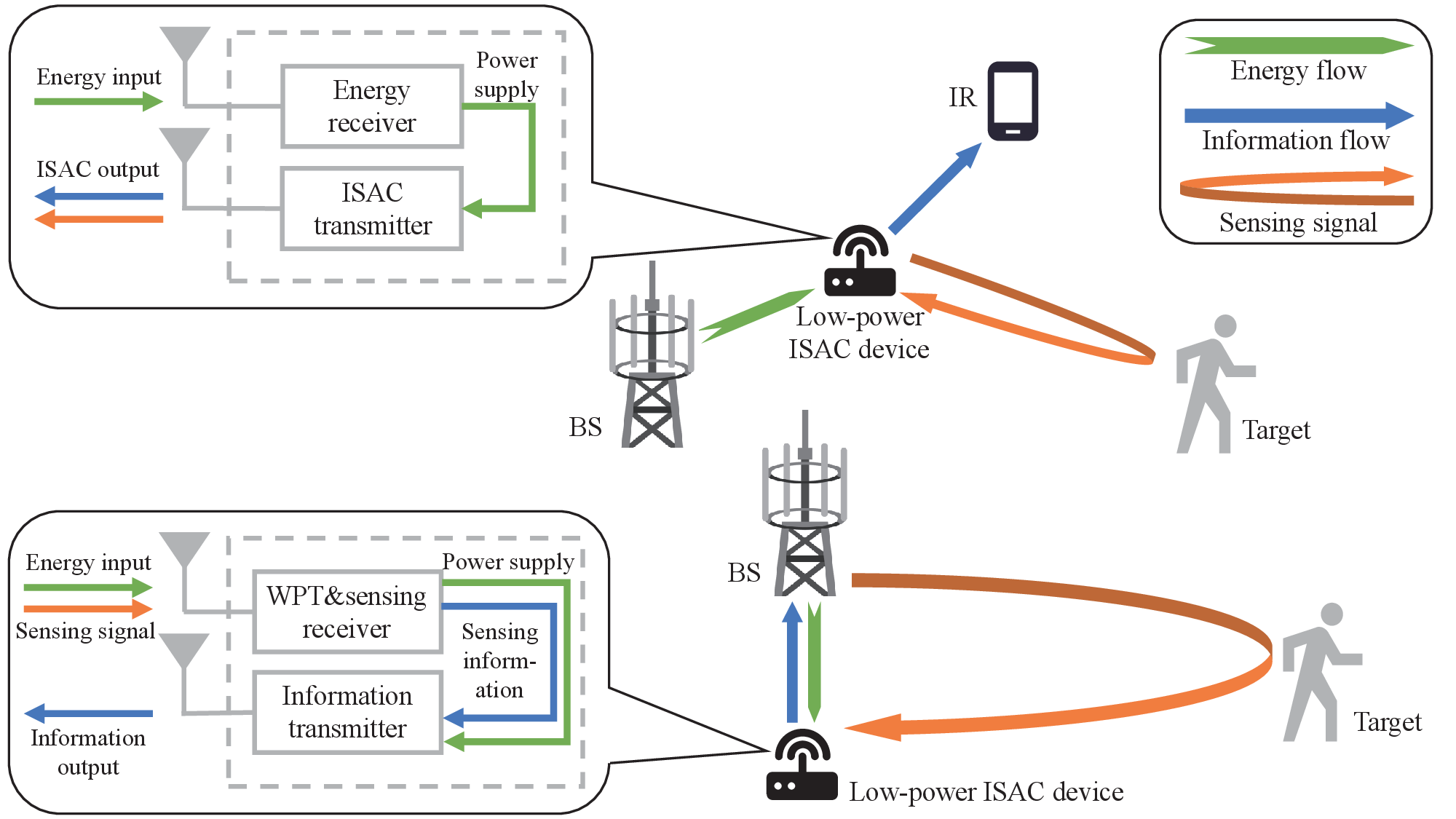}} \caption{Wireless powered ISAC.}
\end{figure}

Another interesting operation mode of the multi-functional ISCAP network is to exploit WPT from BSs or dedicated power beacons (PBs) to supply energy for wireless devices performing ISAC. This mode is particularly appealing for IoT devices that are usually with low power requirements. An experimental demonstration of such a system has been reported in \cite{li2023integrating}. 

In practice, wireless powered ISAC can be realized in various ways. Take the case with one single IoT device as an example. As shown at the top of Fig. 5, the BS implements the WPT to charge the low-power ISAC device in the downlink. Then, the ISAC device leverages the harvested energy from the BS to implement both monostatic sensing and communications. This system can operate using either time-division duplexing (TDD) or FDD protocols, in which the downlink WPT of the BS and the ISAC transmission of the devices are implemented over orthogonal time slots (at the same frequency band) or over orthogonal frequency bands, respectively. For both TDD and FDD operations, new wireless energy harvesting or energy causality constraints are imposed at the wireless powered ISAC device, specifically, the accumulative energy exploited for ISAC at any given time instant cannot exceed the amount harvested from the BS. Subject to such new energy constraints, it becomes necessary to explore novel joint energy-supply-side (for WPT) and demand-side (for ISAC) designs over time to optimize the sensing and communication performance tradeoffs. 

For the aforementioned system, consider a particular transmission period of interest. An interesting problem is to jointly optimize the WPT (e.g., transmit energy beamforming or power allocation) at the BS and ISAC (e.g., joint sensing and communication beamforming) at the low-power device, for maximizing the average communication rate and sensing accuracy over this period, subject to the energy causality constraints. This problem can be particularly challenging to solve. When the wireless channels remain unchanged or the CSI is known \textit{a priori}, this problem is generally solvable via convex and non-convex optimization techniques. Given concave increasing utility functions, it is expected that the BS prefers to allocate more energy at early time slots for WPT, and the IoT device prefers to adopt monotonically increasing power allocation over time to maximize the utility. By contrast, when the wireless channels fluctuate over time or the CSI is unknown, the problem becomes a stochastic optimization problem, for which the dynamic programming or reinforcement learning techniques may be applied \cite{wang2020optimal}.  

The other operation mode is shown at the lower-half of Fig. 5, in which the BS can reuse the energy signals for the dual role of both powering and sensing, and low-power IoT device uses the harvested energy to perform bi-static sensing and data transmission. In particular, both energy harvesting and bi-static sensing at the IoT device can be performed in either time switching or power splitting manners \cite{clerckx2018fundamentals}. In the former approach, the IoT device performs bi-static sensing and energy harvesting over different time instants; while latterly, the IoT device can split the received signal into two portions for bi-static sensing and energy harvesting, respectively. In this mode, the IoT device's energy exploited for sensing signal processing and transmission comes from that harvested from the BS. As such, we need to jointly design the BS's transmission for WPT and sensing as well as the IoT device's energy usage strategy.

When multiple IoT devices are involved, it becomes necessary to design multiple access techniques and coordinate the operation of multiple IoT devices to manage interference. One simple approach is to employ the time division multiple access (TDMA) protocol. In the first slot, the BS can transmit energy signals to multiple IoT devices for energy harvesting simultaneously, and in each of the subsequent time slots, each IoT device can perform ISAC without interfering with each other. More advanced multiple access techniques such as spatial division multiple access (SDMA) and NOMA can also be applied, in which the IoT devices can perform data transmission at the same time, and the inter-user interference can be properly managed via spatial beamforming or successive interference cancellation. The management of joint power supply at the BS and ISAC demand at the IoT devices in this scenario is more important yet challenging.



\section{Multi-BS Cooperation for Multi-functional Networks}

The previous discussion focused on different multi-functional operation modes coordinated by one single BS. Nevertheless, future 6G networks are expected to be densely deployed with the collaboration of massive BSs, in which the inter-BS interference becomes more severe. Therefore, the development of the 6G network expresses the requirement for the reuse of existing infrastructures. For instance, by exploiting the advancements in coordinated multi-point (CoMP) transmission/reception, cloud radio access networks (C-RAN), and cell-free MIMO, densely deployed BSs can be coordinated in signal transmission and reception to realize multi-static and distributed MIMO sensing, enable more efficient interference mitigation and utilization for communication, as well as distributed energy beamforming for WPT. Various new research directions emerge along this networked multi-functional transmission, as described below.

First, consider the networked simultaneous multi-functional transmission, in which multiple BSs need to coordinate their transmission for realizing the three functions simultaneously. This design, however, is rather challenging, as the three functions follow distinct design principles by treating the inter-BS interference differently. In particular, for communications, the inter-BS interference is severely detrimental to performance. However, with perfect data and CSI sharing as well as symbol-level synchronization among BSs, the inter-BS interference turns out to be beneficial to convey useful information through CoMP transmission. For sensing, whether the inter-BS interference is useful highly depends on the synchronization among BSs: If the BSs are perfectly synchronized, then the cross-BS echo signal (i.e., the echo signal from one BS to target to another BS) can be utilized for effective target detection and localization \cite{liu2023seventy, perceptiveNetwork}. For WPT, the inter-BS interference is always beneficial, as it also contains energy that can be harvested by ERs. 

Under these principles, the multi-BS transmission designs for the three objectives are different. Consider a multi-BS setup, in which each BS sends individual messages to one IR, and these BSs cooperatively serve one ER and sense one target. It is clear that the information beam from each BS should steer towards its desired IR, while avoiding other IRs to minimize interference. By contrast, the energy and sensing beams should steer towards the common ER and target, for enhancing the WPT and sensing performances. Therefore, when the three different types of nodes (IR, ER, and target) coexist, sophisticated optimization is generally needed to ensure the performance requirements while minimizing communication interference. The simple example clearly shows the insights in the design of information, sensing, and energy beamformers. The joint design becomes challenging if more advanced joint information transmission and distributed MIMO wireless sensing are implemented, especially when only partial or imperfect CSI is available.

Next, it is also interesting to consider multi-BS cooperation for sensing-assisted WIPT and wireless-powered ISAC. In particular, when there are multiple BSs, they can collaborate in multi-static sensing to acquire detailed environment and scatterers information over a large-scale area, based on which it becomes possible for the BSs to reconstruct the environment and even channel knowledge map \cite{zeng2023tutorial}.
Such an advancement is expected to significantly reduce the signaling overheads for acquiring global CSI for networked WIPT. In this case, developing training-free or light-training protocols is a noteworthy challenge.

On the other hand, for multi-BS enabled wireless powered ISAC, certain BSs can be employed as energy transmitters that perform distributed energy beamforming for maximizing the transferred energy to IoT devices. Meanwhile, some other BSs can be utilized as information fusion centers to collect sensing information from IoT devices. In such scenarios, the role of duplexing techniques becomes increasingly crucial, and the mixed uplink-downlink interference is more complex to handle. How to support network-wide optimization is thus a challenging problem to be tackled in the future.

\section{Future Research Directions}


\begin{figure}[tb]
    \centering {\includegraphics[width=0.45\textwidth]{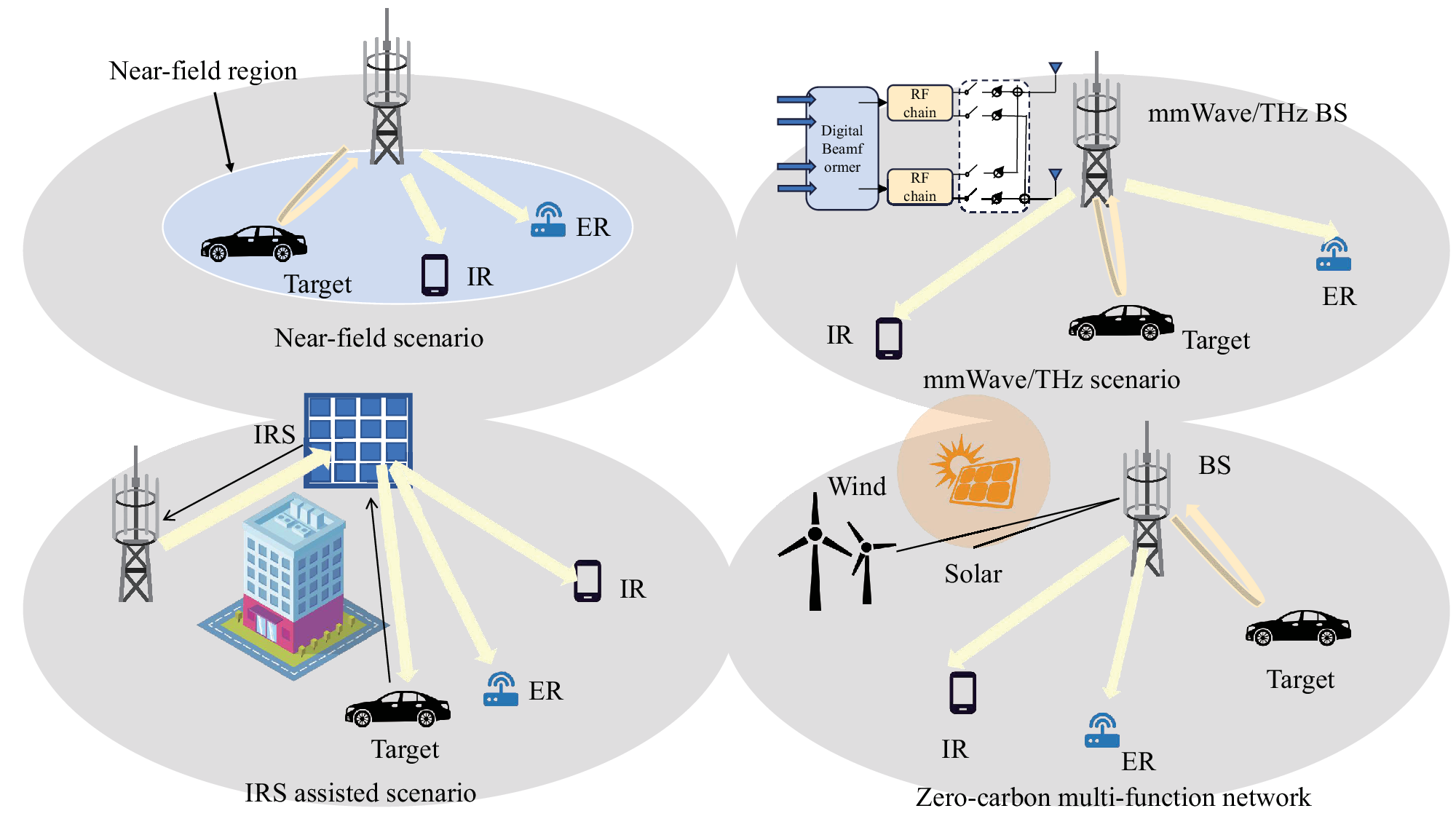}} 
    \caption{New ISCAP scenarios for future research.}
\end{figure}

Despite these advancements, multi-functional wireless networks introduce abundant research opportunities and challenges. In the following, we list some interesting directions to motivate future research as shown in Fig. 6.
\begin{itemize}

\item \textbf{Near-field multi-functional transmission with extremely large-scale antennas:} MIMO is continuously evolving towards extremely large-scale antennas for 6G. In this case, conventional far-field channel models based on planar wavefront become inaccurate, and new near-field channel models relying on spherical wavefront are necessary \cite{lu2023tutorial}. Near-field transmission offers new opportunities. First, extremely large-scale antennas are expected to enable accurate three-dimensional (3D) localization for resolving targets in both angle and distance domains, even for cases with narrowband transmission only. Next, beam focusing makes it possible to concentrate wireless energy at specific physical points with significantly enhanced communication signal-to-noise ratio (SNR) and energy transfer efficiency, which also enables new location-aware interference mitigation and location division multiple access. How to acquire CSI with massive dimensions and design effective transmit strategies for balancing the sensing-communications-powering performance tradeoff becomes an important and worthy pursuit for future research. 

\item \textbf{MmWave/THz multi-functional transmission:} The mmWave and THz spectrums are expected to be widely adopted in 6G at higher frequencies with large bandwidths. MmWave/THz is also a natural fit to be integrated with massive MIMO, which is anticipated to provide extremely high communication data rate, high-precision sensing, and localization, as well as improved energy transfer efficiency (by exploiting the potentially large beamforming gains). To reduce hardware costs, hybrid digital-analog transceiver architectures become widely adopted. For this architecture, the acquisition of CSI becomes more challenging, and sensing assistance may act as a viable solution to tackle this issue. In this case, the choices of hybrid architectures and the design of hybrid beamforming are important problems for enabling multi-functional networks.

\item \textbf{Intelligent reflecting surface (IRS)}: Recently, IRS has emerged as a promising technique to enhance the performance of sensing, communication, and powering. In particular, IRS can reshape the wireless environment to establish reflective virtual LoS links, enhance signal strengths, mitigate harmful interference, and modify the channel rank to facilitate ISAC \cite{10243495} and WIPT \cite{zhang2022near}. The design of joint transmit beamforming at BS and reflective beamforming at IRS for optimizing the C-R-E performance is also worth studying for the simultaneous multi-functional transmission. Furthermore, sensing can be performed to help the IRS acquire its associated CSI, which is normally difficult to obtain in conventional IRS networks. To this end, the deployment of novel transmission protocols becomes necessary. 
 
\item \textbf{Zero-carbon multi-function network with new energy technologies:} As energy consumption and carbon footprint of wireless networks are becoming a major concern for our society, which may become more serious, due to the integration of multiple functions. With the emergence of renewable energy, energy storage, and smart grid, it is possible to realize highly carbon efficient and even zero-carbon multi-function networks, by utilizing such energy technologies as new energy supply. 
On the other hand, the dynamic communication resource allocation is pivotal in effectively matching the random energy arrival and demands of the three functionalities. Therefore, to enhance the system performance for such hybrid-energy-supplied networks, it is essential to jointly design the energy supply side (for energy storage management and energy trading with grids) and the energy demand side (for wireless resource allocation), with the new objective of maximizing the carbon efficiency while ensuring the performance requirements on communications, sensing, and powering. 
\end{itemize}

\section{Conclusion}

This article presented novel multi-functional ISCAP networks integrating WIPT and ISAC. We discussed four operational models, including simultaneous multi-functional transmission, sensing-assisted WIPT, wireless-powered ISAC, and networked multi-functional transmission. For each of these modes, we unveiled new waveform and beamforming designs, novel transmission protocols, innovative joint resource allocation designs, and new interference management and exploitation methods to facilitate the multi-functional operation and enable their mutual assistance. Interesting research directions were also discussed. It is our hope that this article can pave the way towards future self-sustainable 6G networks, which will be crucial for supporting the sensing and communication needs of massive IoT devices.

\bibliographystyle{IEEEtran}
\bibliography{my_ref}

\end{document}